\documentclass[twocolumn]{aastex63}

\begin{document}

\title{The origin of the nuclear star-forming ring in NGC 3182}

\correspondingauthor{Mina Pak}
\email{minapak@kasi.re.kr}

%\author[0000-0002-0786-7307]{Mina Pak}
\author{Mina Pak}
\affiliation{Korea Astronomy and Space Science Institute (KASI), 776 Daeduk-daero, Yuseong-gu, Daejeon 34055, Republic of Korea}

\author{Joon Hyeop Lee}
\affiliation{Korea Astronomy and Space Science Institute (KASI), 776 Daeduk-daero, Yuseong-gu, Daejeon 34055, Republic of Korea}

\author{Hyunjin Jeong}
\affiliation{Korea Astronomy and Space Science Institute (KASI), 776 Daeduk-daero, Yuseong-gu, Daejeon 34055, Republic of Korea}

\author{Woong-Seob Jeong}
\affiliation{Korea Astronomy and Space Science Institute (KASI), 776 Daeduk-daero, Yuseong-gu, Daejeon 34055, Republic of Korea}
\affiliation{Korea University of Science and Technology (UST), 217 Gajeong-ro Yuseong-gu, Daejeon 34113, Republic of Korea}

\begin{abstract}

We investigate the stellar and ionized gas kinematics, and stellar populations of NGC3182 galaxy using integral field spectrograph data from the Calar Alto Legacy Integral Field Area survey. We try to clarify the nature of the ring structure in NGC 3182. We find a negative stellar age gradient out to the ring, while [$\alpha$/Fe] considerably enhanced in the ring. The stellar metallicity shows a smooth negative gradient. From the line ratio diagnostic diagrams, we confirm that NGC 3182 is a Seyfert galaxy from emission line flux ratio, while the gas in the inner ring is ionized mostly by young stars. However, any obvious feature of outflows is not found in its gas kinematics. In the ring, star formation seems to have recently occurred and the gas metallicity is slightly enhanced compared to the center. From our results, we conclude that star formation has occurred in the circumnuclear region within a short period and this may result from a positive feedback by AGN radiation pressure. 

\end{abstract}

%\keywords{editorials, notices --- 
%miscellaneous --- catalogs --- surveys}

\section{Introduction} \label{sec:intro}

Nuclear rings are the configurations of a chain of star-forming regions around galactic nuclei. In observations, they were first identified as multiple spots near galaxy centers (\citealt{Mor58}; \citealt{Ser58}) and are usually found in barred mid-type spiral galaxies \citep{Com10}. In numerical simulations, nuclear rings are believed to form as a result of an inflow of gaseous materials by the non-axisymmetric structures such as spiral arms, elongated bulges, and ovals in disk galaxies (\citealt{Com85}; \citealt{Gar91}; \citealt{Ath92}; \citealt{Kna95}; \citealt{Pat00}; \citealt{Kim12}; \citealt{Ath94}; \citealt{Kim14}; \citealt{Moo21}; \citealt{Moo22}). 

\begin{figure}
\centering
\includegraphics[width=8cm]{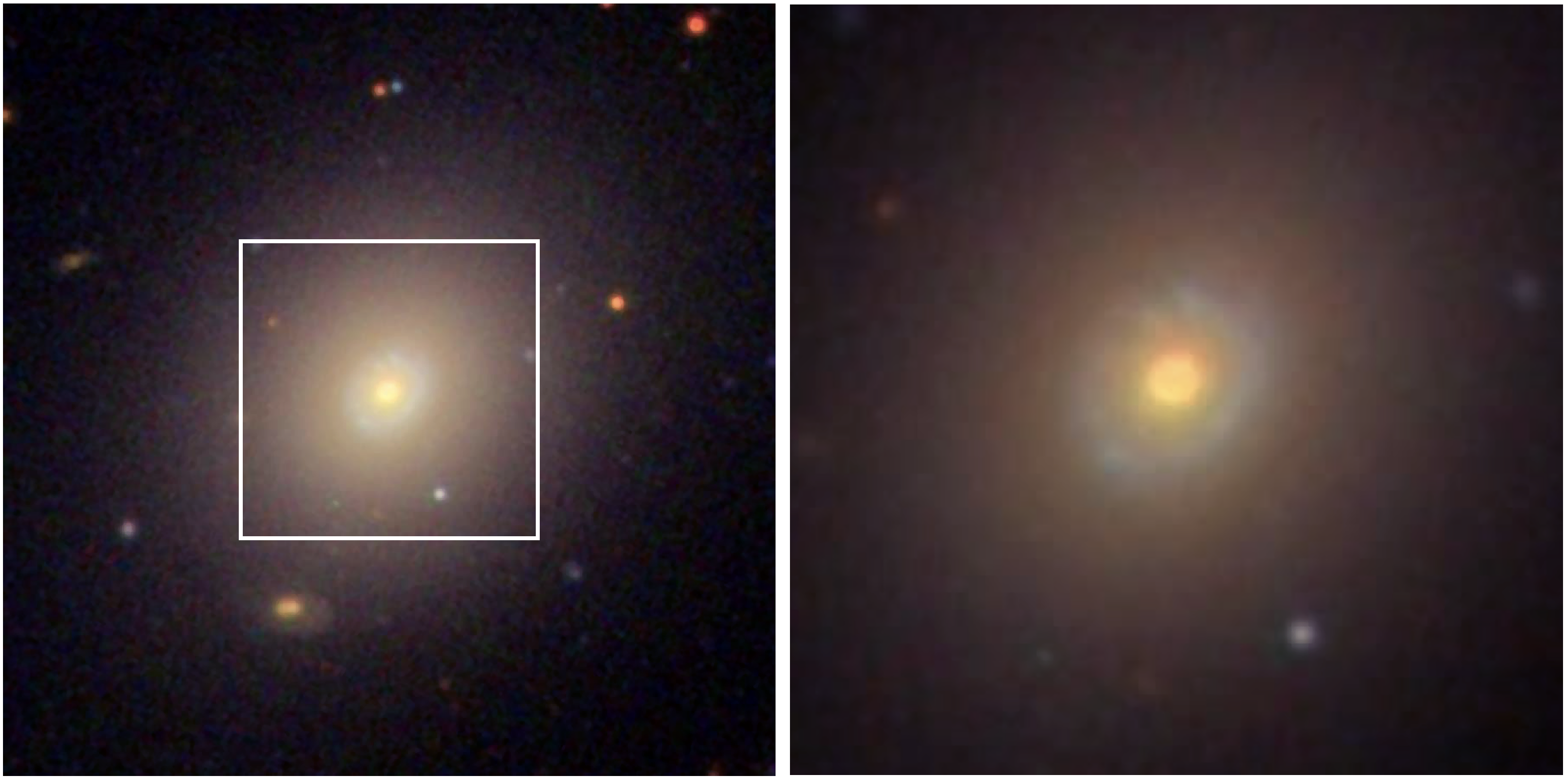}
\caption{In the left, the SDSS g-, r-, and i-bands composite image ($130\arcsec \times 130\arcsec$) of NGC 3182 galaxy. In the right, the zoom-in image of the central $50\arcsec \times 50\arcsec$ (white box in the left column). North is to the top and east is to the left.}
\label{F11}
\end{figure}

However, it is not clear if the bar process is the sole origin of inner rings, because a galaxy with an inner ring but without a bar is found, although it is rare. NGC 3182 is a lenticular (S0) galaxy (SA(r)0+, \citealt{But15}) with stellar mass of $10^{10.3}$ M$_{\odot}$, rather isolated, without any morphological signs of interaction recently. The intriguing feature of NGC 3182 is that it has a nuclear star-forming ring, despite the fact that it shows no evidence of a bar (Figure \ref{F11}). If the inner ring of NGC 3182 did not originate from bar processes, then how did it form? What alternative mechanisms have driven the formation of the nuclear ring in this galaxy?

The importance of active galactic nucleus (AGN) feedback for regulating the star formation for galaxies has been extensively discussed in observations and simulations. Active galactic nuclei (AGNs) are believed to play a critical role in driving the evolution of galaxies releasing an amount of energy by AGN with various physical forms such as radiation, outflows, or radio jets (\citealt{Fab12}; \citealt{Morg17}). AGNs are thought to be associated with both suppression and triggering of star formation (so-called the negative and positive AGN feedbacks, respectively), but their detailed mechanisms are not sufficiently understood yet. AGNs and star formation are frequently found in late-type galaxies, but today it is well known that an early-type galaxy may also have an AGN or current star formation, or even both sometimes (\citealt{Sch07}; \citealt{Ala11}; \citealt{Ala13}).

AGN feedback can have a dramatic effect on gaseous halo in massive elliptical galaxies (\citealt{Cho12}; \citealt{Cio17}; \citealt{Eis17}). \citet{Eis17} have studied the interplay between cooling and AGN feedback in more detail using idealized three-dimensional simulations of elliptical galaxies focused on the multiphase gas. In their models including both wind and radiative feedback, the gas cooling down into the center forms a dense and ring-like disk with a hole of gas in the center of halo. They interpreted that this central hole is produced by the wind feedback, which both heats a central gas and accelerates biconical outflow. \citet{Cio17} have presented two-dimensional hydrodynamical simulations for the evolution of medium–high mass early-type galaxies hosting central massive black holes, which contains accurate and physically consistent radiative and mechanical AGN wind feedback with parsec-scale central resolution. In their more massive models, nuclear outbursts last to the present epoch, with large and frequent fluctuations in nuclear emission and from the gas. Especially, star formation takes place in around a few hundred parsecs in a very short time scale. Their results provide interesting constraints on the effect of AGN feedback in early-type galaxies. Thus, according to numerical studies, AGN feedback seems to be an alternative origin of a star-forming nuclear ring. Then, is NGC 3182 its rare observational example?

The purpose of this work is to investigate the origin of the nuclear star-forming ring in NGC 3182 by comparing the spatially resolved stellar populations, kinematics and emission line properties of NGC 3182 using the Calar Alto Legacy Integral Field Area (CALIFA; \citealt{San12}; \citealt{Hus13}) survey; what triggers the star formation on the ring, and how it is related to the stellar populations in the host galaxy. Since NGC 3182 is a rare early-type galaxy that is overall quiescent but hosts a nuclear star-forming ring, we try to clarify the nature of the ring and its star formation history. 

This paper is organized as follows. Section~2 describes the CALIFA data. The analysis of data is in Section~3. We present our results and discussion in Section~4. Finally, conclusions are given in Sections~5. Throughout the paper we adopt a standard $\Lambda$CDM cosmology with $\Omega_m=0.3$, $\Omega_\Lambda=0.7$, and $H_0=73$\,km\,s$^{-1}$\,Mpc$^{-1}$.

%----------------
%----------------
%NUV-r = 3.7 W2 - W3 = 2.0
%%%%%%%%%%%%%%
\section{Data}
%%%%%%%%%%%%%%
We use the integral field spectrograph (IFS) data cube from the CALIFA survey, which observed the 667 nearby galaxies using the PMAS/PPAK spectrograph the 3.5-m telescope at the Calar Alto observatory (\citealt{Rot05}; \citealt{Kel06}). A field of view the PPAK is $74\arcsec \times 64\arcsec$, consisting of $382$ fibers of $2.7\arcsec$ diameter \citep{Kel06}. The survey observed galaxies with the gratings V500 with a nominal resolution ($\lambda$/$\Delta \lambda$) of $850$ at $5000$ {\AA} (FWHM $\sim 6$ {\AA}) and a wavelength range from $3745$ to $7500$ {\AA}, and V1200 with a better spectral resolution of $1650$ at $4500$ {\AA} (FWHM $\sim 2.7$ {\AA}) and ranging from $3650$ to $4840$ {\AA}. More detailed information about the CALIFA survey is available in the papers of the CALIFA team (\citealt{Wal14}; \citealt{San12}; \citealt{Hus13}; \citealt{Gar15}; and \citealt{San16}). In this study for NGC 3182 galaxy, we use the V500 data cube only.

%%%%%%%%%%%%%%%%%%
\section{Analysis} 
%%%%%%%%%%%%%%%%%%

We adopt two different ways to derive mass-weighted and luminosity-weighted stellar populations. From full spectral fitting using template spectra for a wide range of ages and metallicities, we estimate the mass-weighted stellar  population properties. Mass-weighted properties are appropriate to probe the star formation history hardly dominated by young stars. On the other hand, it is well-known that young massive stars can dominate the luminosity-weighted stellar populations, even when they contribute small amounts to the stellar mass \citep{Tra00}. The stellar populations derived from line strengths provide the luminosity-weighted properties, which can be dominated by young stars and therefore may not reflect the genuine distribution of stellar populations in an old galaxy. Nevertheless, the luminosity-weighted properties are particularly useful for investigating recent star formation activities.

For securing sufficient signal-to-noise ratio (S/N), we conduct spatial binning for the IFS data cube in two different ways. Firstly, we spatially bin the CALIFA data by means of the centroidal Voronoi tessellation algorithm of \citet{Cap03} using the PINGSoft \citep{Ros12} software. The minimum S/N per bin for each galaxy is set to be 5, which is marginal value for measuring kinematics. Secondly, spaxels are binned to be a series of elliptical annuli. Six annuli bins from the center to the $9\arcsec$ radius are generated with $1.\arcsec5$ interval. To secure sufficient S/N, two outermost annuli bins from $9\arcsec$ to $23\arcsec$ are generated with $7\arcsec$ interval.

%%%%%%%%%%%%%%%%%%%%%%%%%%%%%%%%%%
\subsection{Full spectral fitting} 
%%%%%%%%%%%%%%%%%%%%%%%%%%%%%%%%%%
We use the PYTHON version of the penalized pixel fitting code (pPXF; \citealt{Cap04}; \citealt{Cap17}), which allows to estimate the stellar and gas kinematics, stellar populations, and gas emission lines simultaneously. 

We estimate the mass-weighted age and metallicity from the CALIFA data using the $150$ MILES single stellar population (SSP) model templates from \citet{Vaz10}, covering a wide range in ages from $0.06$ to $15.8$ Gyr and metallicities from $-1.71$ to $0.22$. We use the \textsc{log\_rebin} routine provided with the pPXF package in order to match the spectral resolution of the MILES SSP templates to that of the CALIFA data. As described in \citet{Van17}, we first run pPXF for each bin to obtain a precise noise estimate from the residual of the fit and then a second iteration to clip outliers using the \textsc{CLEAN} keyword in pPXF. We extract the stellar velocity and velocity dispersion in a third iteration using a 12th-order additive polynomial. In the final step, with the extracted stellar velocity and velocity dispersion fixed, gas fluxes and kinematics, and the mass-weighted mean stellar age and metallicity are estimated by using a 10th-order multiplicative polynomial \citep{Cap17}. The regularization process is not applied in our analysis.

%%%%%%%%%%%%%%%%%%%%%%%%%%%%%%%%%%%%%%%%%%%%%%%%%%%%%%%%%%%%%%%%
\subsection{Measurement of Lick indices and stellar populations} 
%%%%%%%%%%%%%%%%%%%%%%%%%%%%%%%%%%%%%%%%%%%%%%%%%%%%%%%%%%%%%%%%

A number of absorption lines from the Lick Observatory Image Dissector Scanner (Lick/IDS) system can be used as tools to probe the stellar populations of a galaxy \citep{Wor94}. We quantify the absorption line-strength for H$\beta$, Mgb, Fe5270, and Fe5335, using the code of \citet{Gra08}. We first broaden the CALIFA spectra in each bin to match with the spectral resolution of each index in the Lick system (\citealt{Wor97}; \citealt{Tra98}), as implemented in the routine. The observed H$\beta$ line is a combination of intrinsic emission with intrinsic absorption, so the intrinsic H$\beta$ emission needs to be corrected to obtain the genuine H$\beta$ absorption line-strength. We correct the genuine H$\beta$ absorption line-strength by adding up the difference between the observation and best-fit template as the intrinsic H$\beta$ emission.

To derive luminosity-weighted SSP-equivalent parameters, we use the Lick index grid method. The age, metallicity, and abundance ratio are derived by comparing Lick indices to TMB model grids given by \citet{Tho03}, with iterations between different index versus index planes \citep{Puz05}. The [MgFe]$^\prime$ is a good tracer of metallicity because it is insensitive to [$\alpha$/Fe], and H$\beta$ is an age indicator, which is the least sensitive to [$\alpha$/Fe] among the Balmer lines \citep{Tho03}. Therefore, we use the indices plane of [MgFe]\footnote{[MgFe]$^\prime$ $=$ $\sqrt(Mgb \times (0.72 \times Fe5270 + 0.28 \times Fe5335)).$} versus H$\beta$ to estimate the metallicity and age. We also use the Mgb-$<$Fe$>$\footnote{$<$Fe$> =$ (Fe5270 + Fe5335)/2} plane to estimate [$\alpha$/Fe], since the Mgb is sensitive to [$\alpha$/Fe], and the $<$Fe$>$ is a metallicity indicator. TMB models cover a wide range of age $= 1$ to $15$ Gyr, [Z/H] $= -0.33$ to $0.67$, and [$\alpha$/Fe] $= -0.3$ to $0.5$. To obtain fine accuracy in the measurements, we interpolate the TMB models to grids with $\sim 300,000$ individual models, spanning the interval with $140$ steps in age, $30$ steps both in [Z/H] and [$\alpha$/Fe].

%%%%%%%%%%%%%%%%%%%%%%%%%%%%%%%%
\section{Results and Discussion}
%%%%%%%%%%%%%%%%%%%%%%%%%%%%%%%%
%%%%%%%%%%%%%%%%%%%%%%%%%%%%%%%%
\subsection{Stellar populations}
%%%%%%%%%%%%%%%%%%%%%%%%%%%%%%%%
%----------------
\begin{figure*}
\centering
\includegraphics[width=17cm]{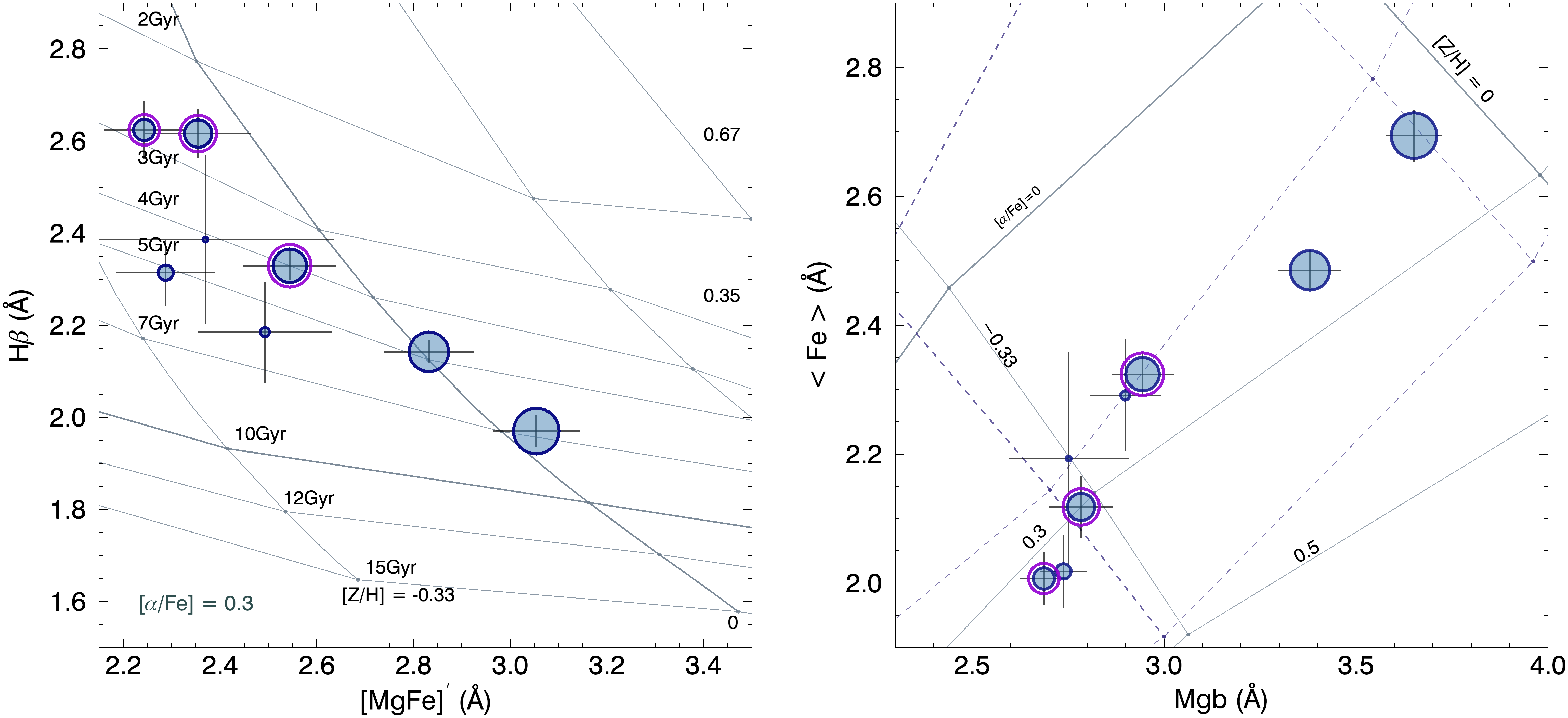}
\caption{H$\beta$ vs. [MgFe]$^{\prime}$ (left) Mg$b$ vs. $<$Fe$>$ (right) diagrams for NGC 3182 from center to outskirt with TMB model grids for ages, metallicities, and abundance ratios. The line strengths are measured along annuli. Symbol sizes are proportional to the radius. The regions of the star-forming ring ($3\arcsec - 8\arcsec$) are enclosed with violet circles. In the Mg$b$ vs. $<$Fe$>$ diagram, solid and dashed lines are the $10$ Gyr and $2$ Gyr models, respectively. The model lines of [Z/H] $= 0$ in the right panel are emphasized as thick lines for each age. Each error bar indicates the measurement uncertainties in each annulus.}
\label{F41}
\end{figure*}
%----------------
%----------------
\begin{figure}
\centering
\includegraphics[width=7cm]{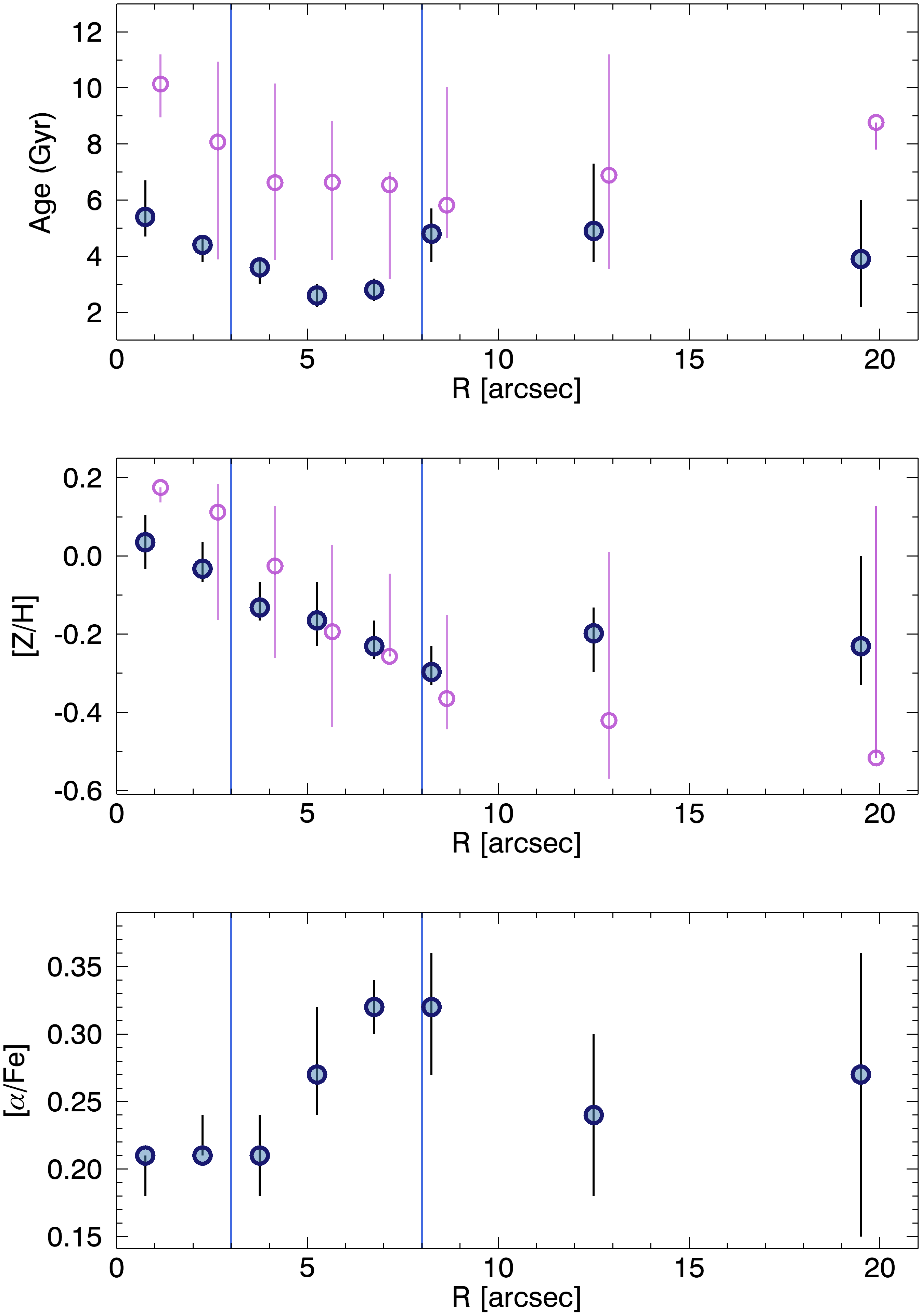}
\caption{Age, metallicity, and $\alpha$-abundance profiles. Filled symbols are from the measurement of lick indices absorption line strengths with the measurement uncertainties on each point are shown. Open symbols are from the measurement of mass-weighted age and metallicity by full spectral fitting using pPXF, the error bars of which connect the maximum and minimum values measured from individual spaxels within each annulus. The vertical lines indicate the inner and outer boundaries of the star-forming ring.}
\label{F42}
\end{figure}
%----------------
%----------------
\begin{figure*}
\centering
\includegraphics[width=13cm]{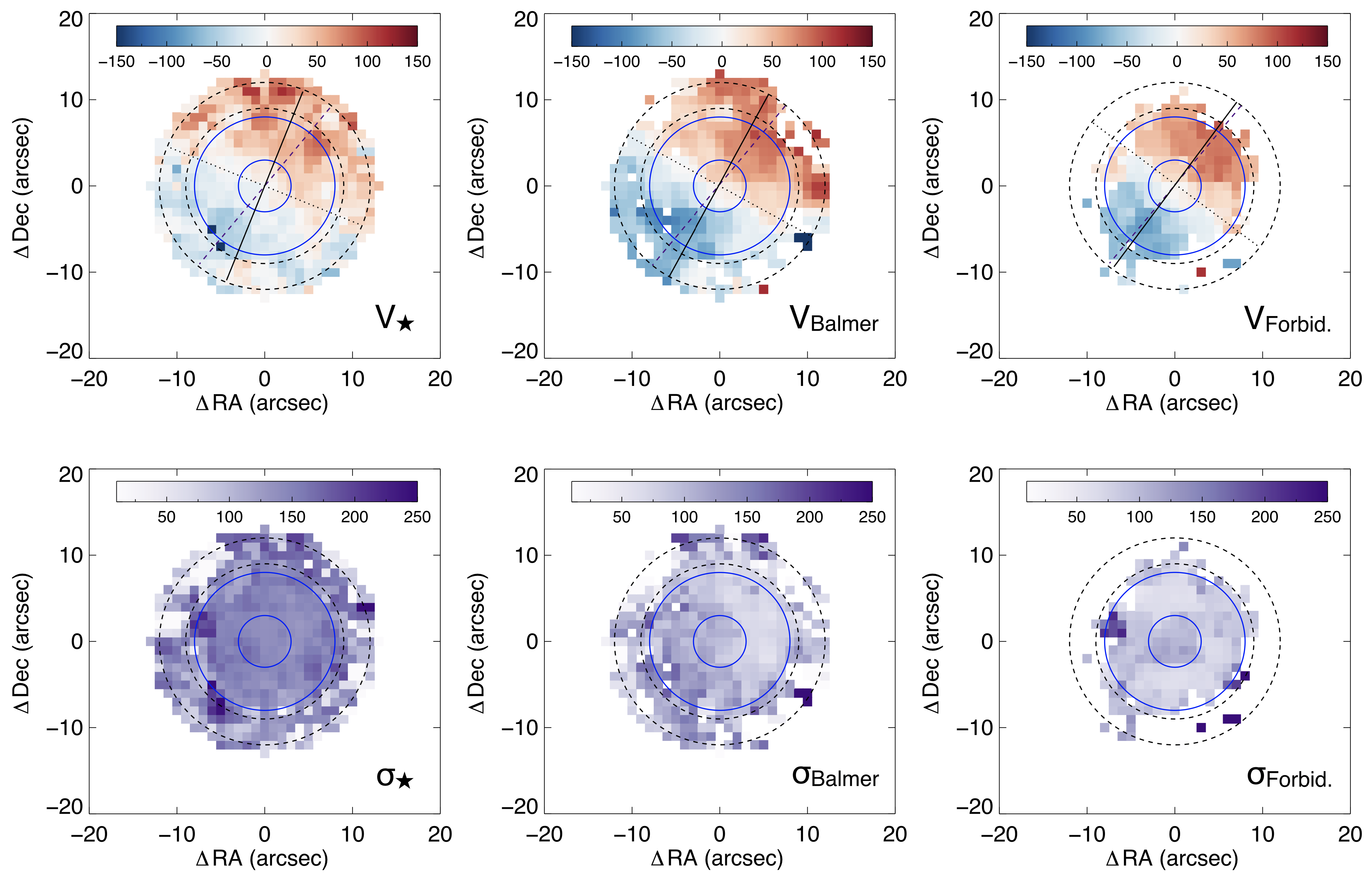}
\caption{The velocity ($v$) and the velocity dispersion ($\sigma$) maps of stellar, balmer, and forbidden lines for the NGC 3182. The solid circles indicate the inner and outer boundaries of the star-forming ring. The inner and outer dashed circles imply the S/N$_{continuum}$ of $\sim30$ and $20$, respectively. H$\alpha > 1.5 \AA$. [NII] or [OIII] $> 1.5\AA$.}
\label{F43}
\end{figure*}
%----------------
Figure \ref{F41} presents index-index diagrams for the selected annuli of NGC 3182 from the measurements of the absorption line strengths. Inner and outer radii of the ring in this study are defined visually using the SDSS optical image and H$\alpha$ intensity map. H$\beta$ clearly increases from the center to the star-forming ring and then decreases again toward outskirts, while the metallicity tends to consistently decrease along radius. These results well agree with \citet{Sil19}. We find that NGC 3182 has around supersolar [$\alpha$/Fe], which means the stars in this galaxy formed in a relatively short time scale. It is particularly notable that [$\alpha$/Fe] is even more enhanced in the star-forming ring. 

The estimates of stellar age, metallicity and $\alpha$-abundance, derived from index measurements show the most striking features in Figure \ref{F42}. The profiles of the age and [$\alpha$/Fe] do not vary significantly, except at the ring radii ($3 - 8\arcsec$). Those profiles indicate that the star formation has recently occurred only in the ring region in a short time-scale. The stellar populations in the star-forming ring are clearly distinct from those in the inner and outer regions of the ring. From this feature, we suspect a possibility that some recent event may be responsible for triggering the ring-shaped star formation in NGC 3182. The age derived from absorption line strength is younger than mass-weighted age measured by full spectral fitting, because H$\beta$ is sensitive to small fractions of recently generated stars.

The metallicity of NGC 3182 shows a smooth negative gradient as usually seen in normal early-type galaxies. This indicate that the source of the recent star formation in NGC 3182 may be the fuel that it originally holds rather than metal-poor fresh gas accretion by merging/interactions with gas-rich neighbor galaxies. Indeed, we do not find any asymmetry or distorted structures in the optical image and kinematics as well, as evidence for any gravitational interactions. Figure \ref{F43} shows the distributions of the stellar and gas velocity and velocity dispersion. NGC 3182 appears to be a regular rotator with the spin-parameter proxy $\lambda$ \citep{Ems11} of stellar component $\sim 0.2$ (at $\epsilon \sim 0.1$). The $\lambda$ of balmer and forbidden lines are $\sim 0.4$ and $\sim 0.5$, respectively. The gas velocities are slightly higher than the stellar velocity. We confirm that there is no evidence of misalignments between the stellar and gas components by measuring of kinematic position angle (PA) using the KINEMETRY method \citep{Kra06}, which quantifies the regularity of the velocity field: $158.0 \pm 7.3$ km s$^{-1}$ from stellar, $151.3 \pm 1.3$ km s$^{-1}$ for balmer lines, and $143.3 \pm 1.8$ km s$^{-1}$ from forbidden lines. This implies that the stellar and gas components are co-rotating in a common structure.

%%%%%%%%%%%%%%%%%%%%%%%%%%%%%%%%%%%%%%%%%%%%%%%%%%%%%%%%%%%%%%%
\subsection{What makes the ring-like star forming structure?} 
%%%%%%%%%%%%%%%%%%%%%%%%%%%%%%%%%%%%%%%%%%%%%%%%%%%%%%%%%%%%%%%
%----------------
\begin{figure*}
\centering
\includegraphics[width=10cm]{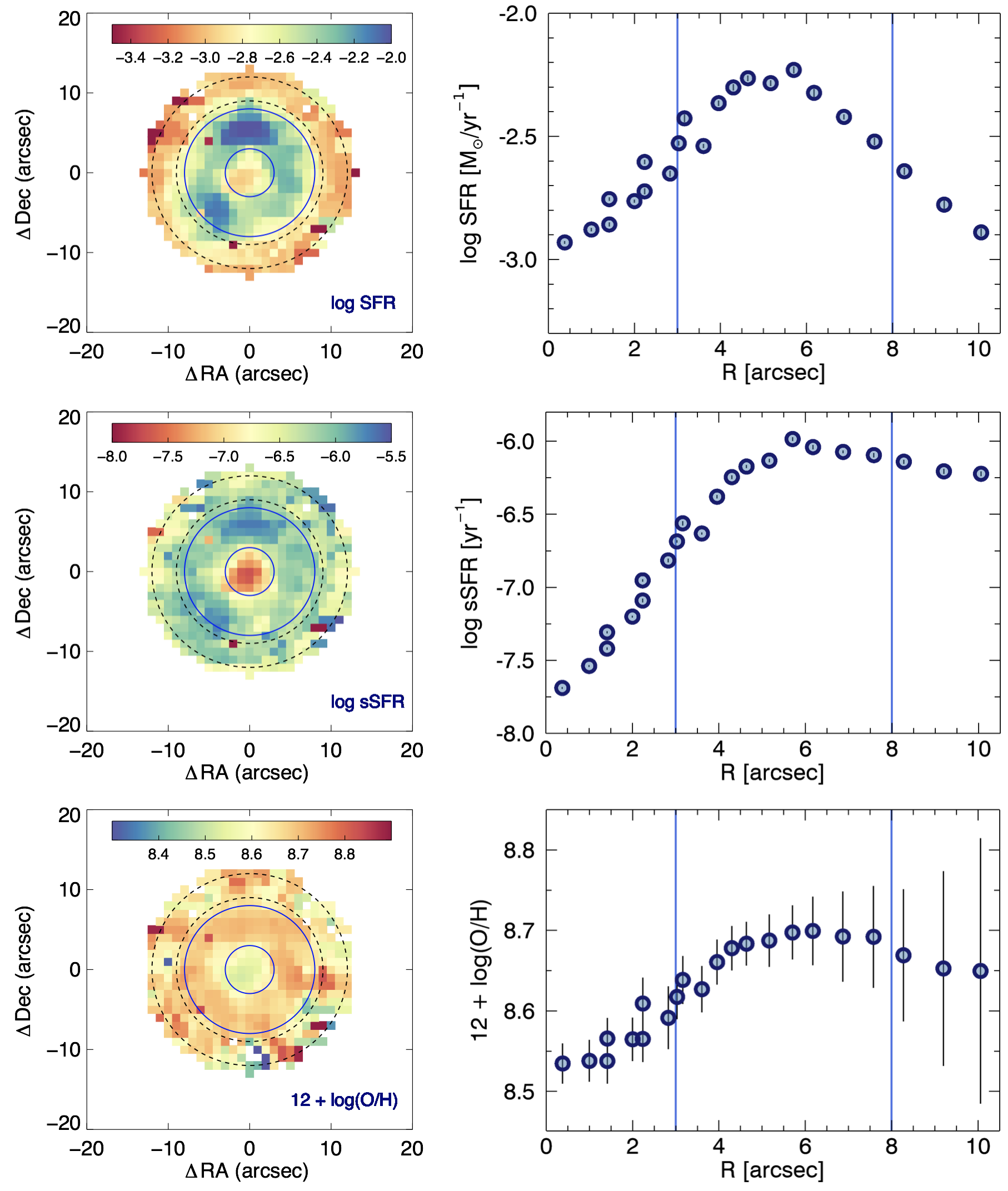}
\caption{The maps (left) and profiles (right) of the specific star formation rate (top) and gas metallicity (bottom) with corresponding uncertainties. The radial profiles are obtained from the isophotal ellipses using the IRAF/ELLIPSE task \citep{Jed87}. The vertical lines distinguish the region of the star-forming ring. The error bars for log SFR and log SFR are shown but their sizes are smaller than the symbol size.}
\label{F44}
\end{figure*}
%----------------

%----------------
\begin{figure*}
\centering
\includegraphics[width=17.5cm]{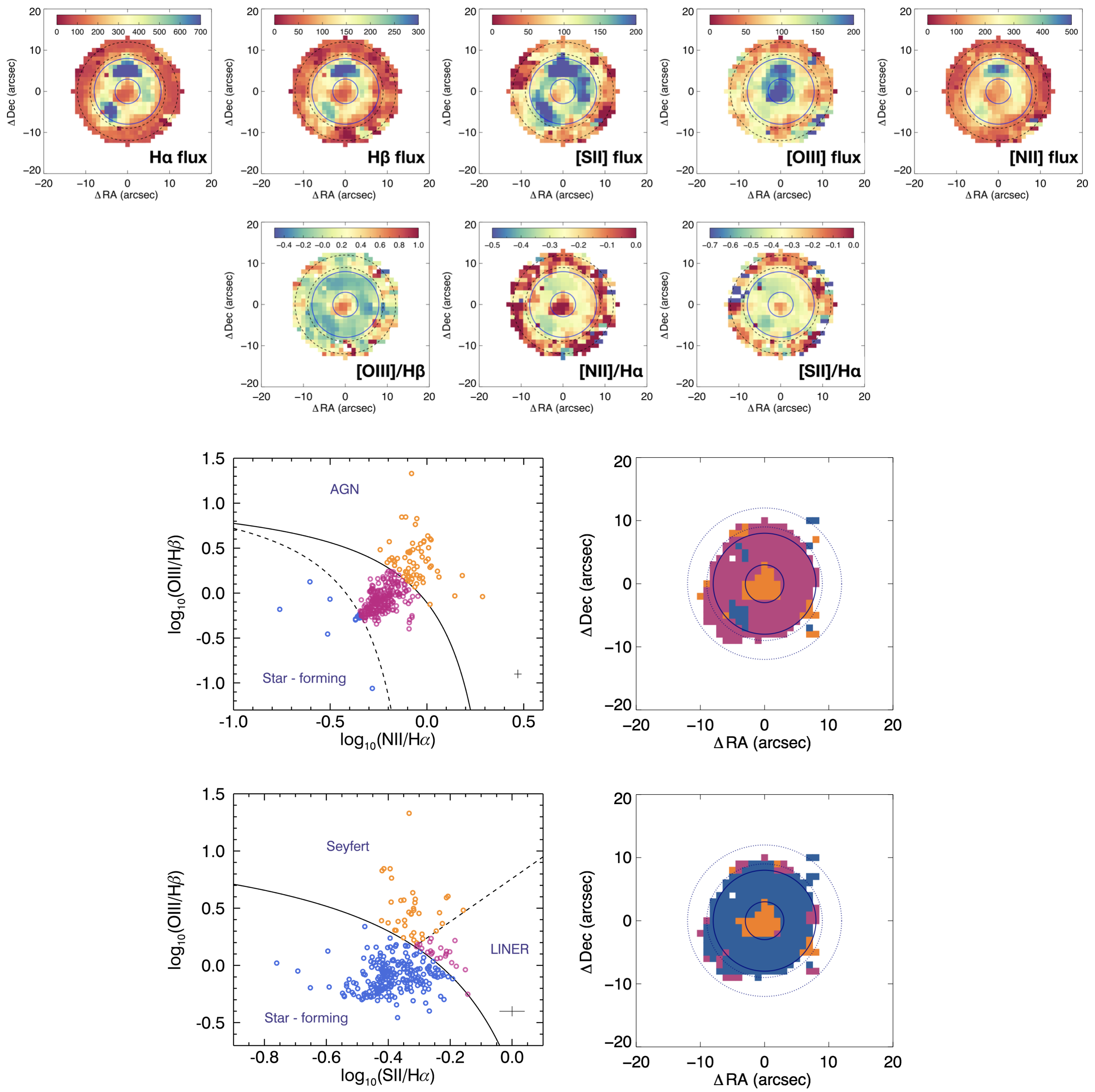}
\caption{In the top, H$\alpha$, H$\beta$, [SII], [OIII] and [NII] line flux ($10^{-17}$ erg s$^{-1}$ cm$^{-2}$) and line-ratio maps of NGC 3182 are presented. In the bottom, we present emission line-ratio diagrams and maps for [OIII]$5007$/H$\beta$ vs. [NII]$6583$/H$\alpha$ and [OIII]$5007$/H$\beta$ vs. [SII]$6731$/H$\alpha$ color-coded by ionization source. The representative uncertainties are shown bottom right corner of each panel. Lines in the diagrams are from \citet{Kau03} and \citet{Sha10}, respectively, to separate star-forming, the AGN, and LINERs. Circles are the same as in Figure \ref{F43}.}
\label{F45}
\end{figure*}
%----------------

A nuclear ring, which is usually associated with the interplay between bar-driven gas inflow and the bar resonances, is a common feature in star-forming barred mid-type spiral galaxies \citep{Com10}. However, the optical image and the IFS-based maps (velocity, velocity dispersion, and line flux) hardly show any clear signs of a bar in NGC 3182. Although it may be somewhat rash to completely exclude the possibility that there exists a very weak bar, at least from our data we found no evidence that NGC 3182 has a bar. Thus, in this paper, we focus on the scenarios other than the bar-driven ring formation.

The star formation from new fresh accreted gas seems to be implausible. Like the stellar metallicity, the high gas metallicity in the top of the Figure \ref{F44} also supports that the fuel source for these flows is highly processed gas, not cold metal-poor gas accreted from outside. We estimated gas metallicity using the following empirical relationship derived by \citet{Pet04}:
\begin{equation}
  12 + log(O/H) = 8.73 - 0.32 \times O3N2.
\end{equation}
The line ratio we used is [OIII]/[NII], referred to as O3N2, initially brought up as an estimation of the metallicity by \citet{All79}. This line ratio metallicity determination was improved using a larger empirical library by \citet{Pet04} and is defined by
\begin{equation}
  O3N2 = log \frac{[OIII]\lambda5007/H\beta}{[NII]\lambda6584/H\alpha}.
\end{equation}
The enhancement of gas-phase metallicity well coincides with the star-forming ring region. Such high gas metallicity in the ring indicates that the star formation may be based on \textit{in situ} materials rather than the fuel from accreted fresh gas. The gas that has originally rich metallicity may have even more enriched by the nucleosynthesis processes during the new star formation.

We estimate the star formation rate (SFR) and the specific star formation rate (sSFR). The SFR is estimated from the H$\alpha$ luminosity as suggested by \citet{Ken98} with a Salpeter initial mass function \citep{Sal55}: % 
\begin{equation}
  SFR~(M_\odot~yr^{-1}) = 7.9 \times 10^{-42}~L_{H\alpha}.
\end{equation}
The H$\alpha$ luminosity, L$_{H\alpha}$ is calculated using
\begin{equation}
  L_{H \alpha} = 4\pi~D^2~F_{H \alpha},
\end{equation}
where D is the cosmological luminosity distance of $33.4$~Mpc taken from the NASA/IPAC Extragalactic Database\footnote{http://ned.ipac.caltech.edu} and F$_{H \alpha}$ is the flux of the H$\alpha$. Then, the estimated SFR is divided by the mass derived from each bin. The galactic extinction from the balmer decrement, E(B - V) $= 0.934 \times$ ln[(F$_{H\alpha}/$F$_{H\beta})/2.86$] is almost zero, thus negligible. In Figure \ref{F44}, it is clearly shown that both the log SFR and the log sSFR reach $-2.23 \pm 0.01$ M$_\odot$~yr${^-1}$ and $-5.99 \pm 0.01$ yr${^-1}$, respectively, within the $3-8 \arcsec$ ring region. The stellar mass density of star-forming ring region spans a range of $5 - 5.5$ M$_\odot$~kpc$^{-2}$. Both the SFR and the sSFR in the ring are slightly enhanced than the star-forming main sequence for resolved pixels in star-forming galaxies \citep{Abd17}.

If such additional star formation in NGC 3182 originated neither from the bar process nor from fresh gas accretion, another possibility is the triggering by AGN outflow. We present two emission line-ratio diagnostic diagrams for [OIII]5007/H$\beta$ versus [NII]6583/H$\alpha$ \citep{Kau03} and the [OIII]5007/H$\beta$ versus [SII]6731/H$\alpha$ (\citealt{Bal81}). Both diagrams reveal that the ionization source in the central region is a AGN/Seyfert (inner $\sim3 \arcsec \sim0.5$kpc). On the other hand, the [OIII]5007/H$\beta$ versus [SII]6731/H$\alpha$ diagram shows that the emission-line ratios are consistent with the photoionization by young stars around the central region ($\sim 3 - 8 \arcsec \sim 0.5 - 1.3$kpc). The [OIII]5007/H$\beta$ versus [NII]6583/H$\alpha$ diagram classifies the ionized source in this region as transition. From the results, we confirm that NGC 3182 holds an AGN in its center but it may not be strong enough since the bipolar galactic-scale outflows of the ionized gas is obviously invisible (Figure \ref{F43}). 

The hydrodynamical simulation of \citet{Cio17} showed that AGN-driven outflows in early-type galaxies can enhance in-situ star formation at the central few hundreds pc radius within a very short time scale of $\lesssim 0.1$ Gyrs (see Figure 7 in \citealt{Cio17}). Star formation is very active in the cold filaments close to the center during the outburst, and one can see a lower SFR region at the fading outburst close to the nucleus due to AGN wind. Even though the stellar mass of NGC 3182 is much less than their massive model, this feature nicely agrees with the NGC 3182. 

Even though the information from \citet{Cio17} is insufficient for precise comparison with our observational results, we tried to check whether their models are plausible for NGC 3182. Figure 7 of \citet{Cio17} shows a star-forming ring structure around an AGN, which may be a fine model for the center of NGC 3182. Under the assumption of a spherical shape, SFR density (M$_\odot$~yr$^{-1}$~kpc$^{-3}$) is estimated to be about 0.06 in the ring of NGC 3182. Although the SFR density in Figure 7 of \citet{Cio17} is simply presented by a color bar, not as exact values, this value seems to be in a similar order to the SFR density of the star-forming region (see the second panel of Figure 7 of \citealt{Cio17}). However, since we do not know the actual three-dimensional shape of the star-forming ring in NGC 3182, the estimated SFR density may have a large uncertainty. In a qualitative regard, the AGN may have caused the ring formation, as presented in the models from \citet{Cio17}, but more precise comparisons between the models and the observations will be necessary for better constraints on the physical conditions of the center of NGC 3182.

This enhancement may originate from the compression of molecular gas clouds along the galactic-scale outflows possibly caused by the AGN. CO is detected in the center of NGC 3182 from the Combined Array for Research in Millimeter Astronomy (CARMA) CO imaging \citep{Ala13}. Though \citet{Ala13} refer that NGC 3182 is one of the faintest detections within the CARMA-ATLAS3D sample, the dust ring is clearly seen in the center of the galaxy. Thus, as the origin of the nuclear star-forming ring in NGC 3182, we suggest a scenario that the key driver of the ring-shaped star formation may be the positive feedback by the AGN. As AGN outflows make a galaxy's gas reservoir condense, the SFR and the gas phase metallicity increases, in particular at the center causing the nuclear star-forming ring structure.

%%%%%%%%%%%%%%%%%%%%%%%%%%%%%%%%%
\section{Summary and Conclusions}
%%%%%%%%%%%%%%%%%%%%%%%%%%%%%%%%%

We have investigated the spatially resolved stellar populations, kinematics and ionized gas properties of NGC 3182 galaxy in the CALIFA survey using full spectral fitting and absorption line population indicators. The SFR and the sSFR estimated from H$\alpha$ flux shows some enhancement in the nuclear ring of NGC 3182. From the stellar age and [$\alpha$/Fe] gradient from index measurements, we conclude that abrupt star formation activity has occurred recently at the ring in NGC 3182. We do not see any evidence of gravitational interactions from its symmetry of stellar and gas kinematics, and thus the nuclear star formation may not be triggered by the accretion of fresh extragalactic gas. Since there is no evidence of a bar structure, the star-forming ring of NGC 3182 may have an origin different from those of late-type galaxies.

We confirm that NGC 3182 hosts an AGN at the center surrounded by star-forming regions from emission line-ratio diagrams. The gas in the ring of NGC 3182 appears to be ionized by young stars and its metallicity is even higher than the gas metallicity at the galaxy center. Such enhancement of gas metallicity may be a natural result of additional star formation from the in-situ gas, possibly triggered by an AGN feedback. Although there is no evidence of kinetic feedback by strong outflows in NGC 3182, a luminous AGN is known to cause symmetric outflows by radiation pressure. Therefore, as the most plausible scenario that can be deduced from our observational analyses, we conclude that the nuclear star-forming ring in NGC 3182 may be formed by a positive feedback from the AGN, through its radiation pressure.

\acknowledgments
We gratefully thank the anonymous referee for constructive comments that have significantly improved this manuscript. MP and JHL acknowledge support from the National Research Foundation of Korea (NRF) grants funded by the Korea government (MSIT) (No. 2022R1C1C2006540 and No. 2022R1A2C1004025, respectively). HJ acknowledges support from the National Research Foundation of Korea (NRF) grant funded by the Korea government (MSIT) (No. NRF-2019R1F1A1041086). This research was supported by the Korea Astronomy and Space Science Institute under the R\&D program (Projects No. 2022-1-850-06 and No. 2022-1-830-06) supervised by the Ministry of Science and ICT. This study uses data provided by the Calar Alto Legacy Integral Field Area (CALIFA) survey (http://califa.caha.es/). Based on observations collected at the Centro Astron\'omico Hispano Alem\'an (CAHA) at Calar Alto, operated jointly by the Max-Planck-Institut f\"ur Astronomie and the Instituto de Astrof\'isica de Andaluc\'ia (CSIC).

%\vspace{5mm}
%\facilities{HST(STIS), Swift(XRT and UVOT), AAVSO, CTIO:1.3m, CTIO:1.5m,CXO}

%\software{astropy \citep{2013A&A...558A..33A},  
%          Cloudy \citep{2013RMxAA..49..137F}, 
%          SExtractor \citep{1996A&AS..117..393B}         }





\end{document}